\documentclass[10pt, journal]{IEEEtran}

\usepackage{stmaryrd}
\usepackage{mathrsfs}
\usepackage{color}
\usepackage[final]{graphicx}
\usepackage{epstopdf}
\usepackage{multirow}
\usepackage{amsmath}
\usepackage{amssymb,times}
\usepackage{cite}
\usepackage{algorithm}
\usepackage{algorithmic}
\usepackage{stfloats}
\usepackage{subfigure}
\usepackage{wasysym}
\usepackage{hyperref}

\newcommand{\ba}{\begin{array}}
\newcommand{\ea}{\end{array}}
\newcommand{\be}{\begin{displaymath}}
\newcommand{\ee}{\end{displaymath}}
\newcommand{\ben}{\begin{equation}}
\newcommand{\een}{\end{equation}}
\newcommand{\bena}{\begin{eqnarray}}
\newcommand{\eena}{\end{eqnarray}}
\newcommand{\beqa}{\begin{eqnarray*}}
\newcommand{\enqa}{\end{eqnarray*}}

\newcommand{\bc}{\begin{center}}
\newcommand{\ec}{\end{center}}
\newcommand{\bi}{\begin{itemize}}
\newcommand{\ei}{\end{itemize}}
\newcommand{\benu}{\begin{enumerate}}
\newcommand{\eenu}{\end{enumerate}}
\newcommand{\bdes}{\begin{description}}
\newcommand{\edes}{\end{description}}
\newcommand{\bt}{\begin{tabular}}
\newcommand{\et}{\end{tabular}}

\newcommand \gammabf{\boldsymbol{\gamma}}

\newcommand \etabf{\boldsymbol{\eta}}

\newcommand \lambdabf{\boldsymbol{\lambda}}

\newcommand \pbf{{\bf p}}

\newcommand \sbf{{\bf s}}

\newcommand \ubf{{\bf u}}

\newcommand \Pbf{{\bf P}}

\newcommand{\Ccal}{\mathcal{C}}

\newcommand{\Ncal}{\mathcal{N}}
\newcommand{\Ocal}{\mathcal{O}}

\newcommand{\circlambda}{\mbox{$\Lambda$
             \kern-.85em\raise1.5ex
             \hbox{$\scriptstyle{\circ}$}}\,}

\IEEEoverridecommandlockouts

\begin{document}

\title{Power Allocation for Coexisting Multicarrier Radar and
Communication Systems in Cluttered Environments
\thanks{This work was supported in part by the National Science Foundation
under grants ECCS-1609393 and ECCS-1923739.}
}

\author{Fangzhou Wang and Hongbin Li,~\IEEEmembership{Fellow,~IEEE}

\thanks{Fangzhou Wang and Hongbin Li are with the
Department of Electrical and Computer
Engineering, Stevens Institute of Technology, Hoboken, NJ
07030
USA (fwang11@stevens.edu; hli@stevens.edu).}

}

\maketitle

\begin{abstract}
In this paper, power allocation is examined for the coexistence of a radar and a communication system that employ multicarrier waveforms. We propose two designs for the considered spectrum sharing problem by maximizing the output signal-to-interference-plus-noise ratio (SINR) at the radar receiver while maintaining certain communication throughput and power constraints. The first is a joint design where the subchannel powers of both the radar and communication systems are jointly optimized. Since the resulting problem is highly nonconvex, we introduce a reformulation by combining the power variables of both systems into a single stacked variable, which allows us to bypass a conventional computationally intensive alternating optimization procedure. The resulting problem is then solved via a quadratic transform method along with a sequential convex programming (SCP) technique. The second is a unilateral design which optimizes the radar transmission power with fixed communication power. The unilateral design is suitable for cases where the communication system pre-exists while the radar occasionally joins the channel as a secondary user. The problem is solved by a Taylor expansion based iterative SCP procedure. Numerical results are presented to demonstrate the effectiveness of the proposed joint and unilateral designs in comparison with a subcarrier allocation based method.
\end{abstract}


\section{Introduction}
\label{sec:introduction}
In recent years, radar has been utilized to numerous civilian applications such as traffic control, remote sensing, car cruise control, and collision avoidance. On the other hand, there is a tremendous demand for additional bandwidth from the wireless sector. Thus, the coexistence between radar and communication systems using shared spectrum has attracted significant attention (e.g.
\cite{ZhengLopsEldar2019,AubryCarotenuto2016,LiPetropulu2015ICASSP,BicaKoivune16ICASSP,LiuMasouros18,WangLiGovoni2019,KangMongaRangaswamy19}). The coexistence, if improperly implemented, can cause
significant interference and performance degradation for both systems
\cite{ShajaiahClancy15,NartasilpaErricolo16,ChiriyathBliss16}. Extensive research has been directed to employing various signal processing techniques, i.e., interference mitigation, power and/or subcarrier allocation, precoding, and waveform design, to allow both radar and communication systems to efficiently share the spectrum.

Depending on the number of systems involved in the design, the
research can be classified into two types. The first is based on
\emph{joint design} that addresses the radar and communication
coexistence by jointly optimizing performance metrics and simultaneously adjusting parameters for both systems \cite{LiPetropulu2017,ZhengLopsWangTSP2018,ChengLiaoHe19,WangLi2019}, e.g., the throughput for the communication system and the signal-to-interference-plus-noise ratio (SINR) for the radar. Specifically, \cite{LiPetropulu2017} considered the co-design of the communication transmit covariance matrix and radar sampling scheme for multiple-input multiple-output (MIMO) radar and communication systems, where the design was formulated as a nonconvex problem that was solved via an alternating optimization algorithm. In \cite{ZhengLopsWangTSP2018}, the radar pulse and communication encoding matrix were jointly designed for the coexistence of a communication system and a pulsed radar. The coexistence of MIMO radar and downlink multiple-input signle-output (MISO) communication systems was
considered in \cite{ChengLiaoHe19} by minimizing the Cram\'er-Rao bound of direction-of-arrival estimation while imposing an appropriate constraint on the communication quality of service. In addition, \cite{WangLi2019} studied the joint design problem of radar and communication coexistence by considering both radar-centric and communication-centric formulations.

The second type is based on \emph{unilateral design} from either the radar or communication perspective \cite{DengHimed13,DengHimed15,ZhengLopsWang2018,ShiSellathurai2018,AubryMaio2015,AubryCarotenuto2016spl}, i.e., only parameters of one system are adjusted by using the information of the other system. One standard approach for the unilateral design is to mitigate mutual interference via either signal processing \cite{DengHimed13,DengHimed15,ZhengLopsWang2018} or constrained optimization \cite{ShiSellathurai2018,AubryMaio2015,AubryCarotenuto2016spl} techniques. Specifically, \cite{DengHimed13} employed receiving beamforming at the radar to cancel the sidelobe interference, while \cite{DengHimed15} proposed a spatial filtering technique to mitigate the wireless interference in both mainlobe and sidelobe directions in coherent MIMO radar. An uncoordinated radar and communication coexistence scenario was considered
in \cite{ZhengLopsWang2018}, where compressed sensing based radar parameter estimation was used in the communication demodulation process to remove the radar interference. Meanwhile, \cite{ShiSellathurai2018} proposed a unilateral design scheme by minimizing the total radar transmission power for an orthogonal frequency division multiplexing (OFDM) radar. Radar waveform design using constrained optimization
techniques to control the amount of radar-to-communication
interference and other waveform properties was investigated in \cite{AubryMaio2015,AubryCarotenuto2016spl}.

Given the wide use of multicarrier signals in communication systems, multicarrier waveforms have become increasingly popular in radar as well due to several advantages such as frequency diversity, waveform diversity, and easy implementation \cite{SenNehorai2011,BicKoivunen2016}. At any time instant, since the desired subcarriers can be digitally selected at the transmitter, narrowband jamming/interferences mitigation can be achieved by simply turning off affected subcarriers. OFDM waveforms with pulse-to-pulse agility was investigated in \cite{LellouchGenderen2008} for Doppler processing from the radar point of view. In \cite{TanBlunt2016}, a sparse spectrum allocation algorithm for an OFDM-type radar was presented by using the integrate sidelobe level (ISL) as an optimization metric. Multicarrier waveforms were also employed by radar and communication systems to tackle coexistence applications \cite{BicaKoivunen2019}.

We consider spectrum sharing between a multicarrier radar and a communication system operating in a cluttered environment, where the communication or radar receiver observes not only the cross-interference from its counterpart but also a multi-path or clutter signal, which arises from the system's own transmission. While multi-path can be exploited for communication, clutter is a self-interference to the radar system and must be adequately mitigated in order to expose weak targets. Clutter is also a \emph{signal-dependent} interference, i.e., it depends on the transmitted signal to be determined, which makes the design problem considerably more challenging \cite{AubryMaioStoica14,QianLopsZheng18,GrossiLops20}. It is noted that multi-path and clutter were neglected in earlier multicarrier-based spectrum sharing studies to ease the development of proposed solutions (e.g., \cite{BicaKoivune16ICASSP,WangLiGovoni2019,BicaKoivunen2019})

Specifically, we propose a joint design approach to jointly optimize the radar and communication transmission power allocated to each subcarrier. The optimum power allocation strategies are obtained for both systems by maximizing the radar output SINR while maintaining a minimum communication throughput constraint, along with a total transmission power constraint and subchannel peak power constraints for each system. The joint power allocation problem is highly nonconvex with respect to (w.r.t.) the design variables. To address this challenge, we reformulate the problem by combining the radar and communication power variables into a single stacked variable. This allows us to bypass a conventional alternating optimization procedure, which is computationally intensive. The resulting problem is then solved by using a quadratic transform method along with a sequential convex programming (SCP) technique.

In addition, we also propose a unilateral design from the radar perspective for the case when the communication system is a primary user of the frequency band, while the radar joins occasionally as a secondary user. The unilateral design optimizes the radar transmission power with throughput and power constraints under the condition that the communication transmission power is a prior knowledge and fixed \cite{ShiSellathurai2018}. The communication system employs a waterfilling solution to allocate subchannel power based on its channel condition when radar is absent. The unilateral design is solved by a Taylor expansion based iterative SCP procedure. Simulation results validate the effectiveness of the proposed joint and unilateral designs over a subcarrier-allocation based method.

The remainder of the paper is organized as follows. The signal model and problem of interest are introduced in \ref{sec:system_model}. The proposed designs along with their solutions are developed in Section \ref{sec:proposed_approach}. Section \ref{sec:simulationresults} contains numerical results and discussions, followed by conclusions in Section \ref{sec:conclusion}.

\emph{Notations}: We use boldface symbols for vectors (lower case) and matrices (upper case). $(\cdot)^T$ the transpose, $\mathbb{E}\{\cdot\}$
represents the statistical expectation, and $\mathcal{O}(\cdot)$ denotes the Landau notation for complexity.
\section{Signal Model}
\label{sec:system_model}
\begin{figure}
\centering
\includegraphics[width=3.1in]{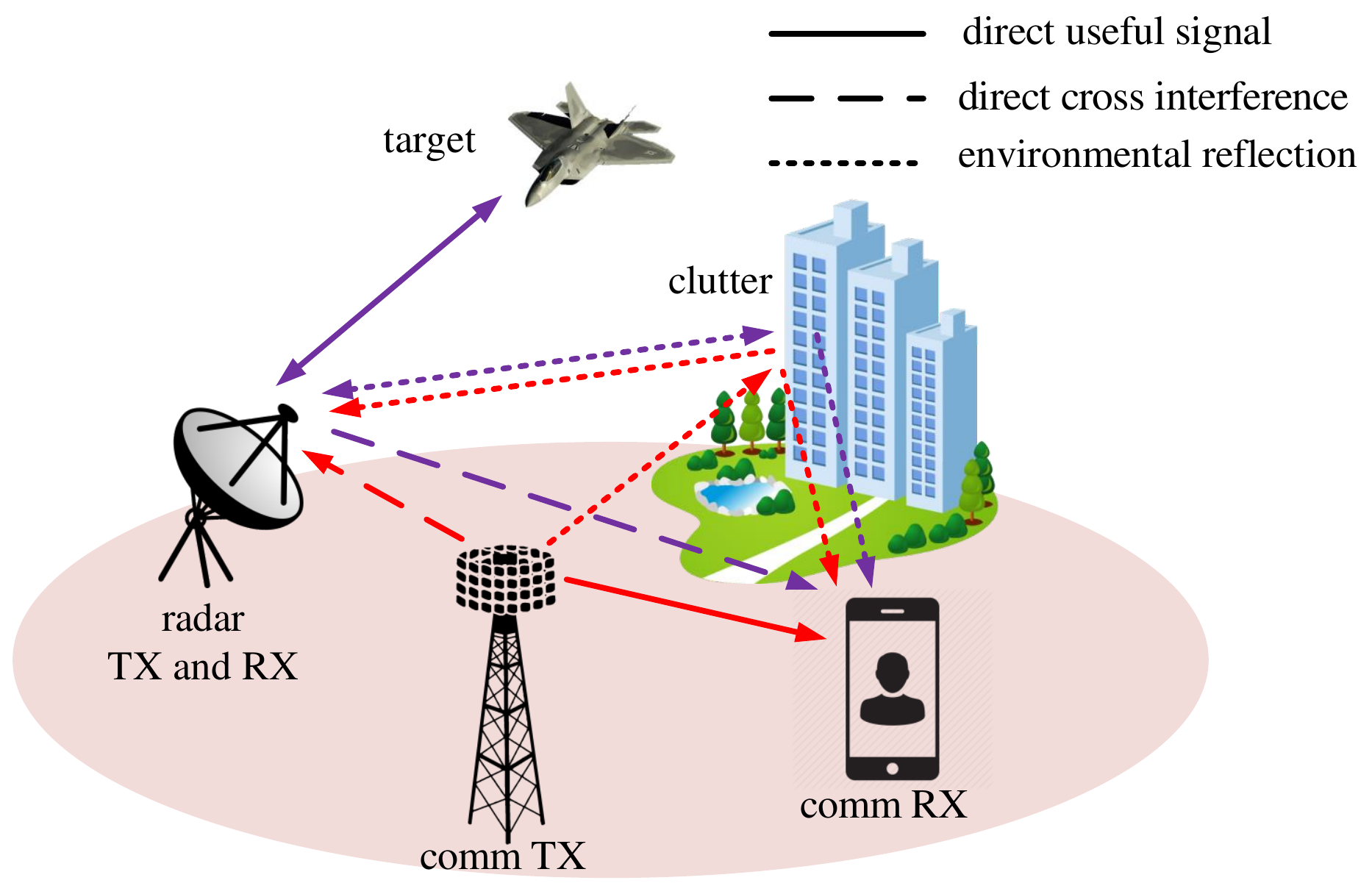}
\caption{A radar and communication coexistence scenario in a cluttered environment.}
\label{fig:configuration}
\end{figure}
Consider a radar system that coexists with a communication system in a cluttered environment as depicted in Fig.\,\ref{fig:configuration}. Both systems share a frequency band of bandwidth $B$ Hz and employ multicarrier waveforms with $N$ subcarriers, where the subcarrier spacing $\Delta f=B/N$. Under the considered set-up, the communication or radar receiver (RX) receives not only the direct useful signal (indicated by the solid lines in Fig.\,\ref{fig:configuration}), but the direct cross interference (dashed lines in Fig.\,\ref{fig:configuration}) and reflections from the environment (dotted lines in Fig.\,\ref{fig:configuration}) as well.

Let $\pbf_\text{c}=[p_{\text{c},1},\cdots,p_{\text{c},N}]^T$ denote the communication powers allocated to the $N$ subcarriers, which are to be determined. Then, the transmitted communication signal can be represented as
\ben\label{equ:transmission_c}
x_\text{c}(t)=q_\text{c}(t)\sum_{n=1}^Nd_n\sqrt{p_{\text{c},n}}e^{j2\pi (f_\text{c}+n\Delta f)t}\triangleq\sum_{n=1}^Nx_{\text{c},n}(t),
\een
where $f_\text{c}$ is the carrier frequency, $q_{\text{c}}(t)$ the communication waveform with a duration $T_{\text{c}}$, and $d_n$ the symbol carried by the $n$-th subcarrier. Without loss of generality, we assume $E\{\vert d_n\vert^2\}=1$.

The multicarrier radar system uses the same carrier frequency $f_\text{c}$ and intercarrier spacing $\Delta f$, and a radar waveform $q_{\text{r}}(t)$ with a duration $T_\text{r}$. For simplicity, we assume $T_\text{r}=T_\text{c}=T$.  Then, the transmitted radar signal can be written as
\ben\label{equ:transmission_r}
x_\text{r}(t)=q_\text{r}(t)\sum_{n=1}^{N}\sqrt{p_{\text{r},n}}e^{j2\pi (f_\text{c}+n\Delta f)t}\triangleq\sum_{n=1}^Nx_{\text{r},n}(t),
\een
where $\pbf_\text{r}=[p_{\text{r},1},\cdots,p_{\text{r},N}]^T$ denote the radar powers that are to be determined.

As illustrated in Fig.\,\ref{fig:configuration}, the signal received at the communication RX on the $n$-th subcarrier is given by\footnote{The channel coefficient are represented by using the following convention. $\alpha'$ denotes a desired (e.g., communication-to-communication or radar-to-radar) channel while $\beta'$ denotes an interference (e.g., radar-to-communication) channel. The subscripts ``cc'' (or ``rc'') indicate the channel starts from the communication (or radar) TX and ends in the communication RX.}
\begin{align}\label{equ:y_c1s}
y_{\text{c},n}(t)&=\sum_{k=1}^{K_{\text{cc}}}\alpha'_{\text{cc},n,k}x_{\text{c},n}(t-\tau_{\text{cc},k})\notag\\&+\sum_{k=1}^{K_{\text{rc}}}\beta'_{\text{rc},n,k}x_{\text{r},n}(t-\widetilde{\tau}_{\text{rc},k})+w'_{\text{c},n}(t),
\end{align}
where $\alpha'_{\text{cc},n,k}$ is the channel coefficient of the $k$-th communication path with propagation delay $\tau_{\text{cc},k}$, $K_{\text{cc}}$ denotes the total number of communication paths, $\beta'_{\text{rc},n,k}$ is the channel coefficient from the radar transmitter (TX) to the communication RX due to the $k$-th clutter scatterer with propagation delay $\widetilde{\tau}_{\text{rc},k}$, $K_{\text{rc}}$ denotes the total number of clutter scatterers, and $w'_{\text{c},n}(t)$ is the additive channel noise. Note that the channel coefficients $\alpha'_{\text{cc},n,k}$ and $\beta'_{\text{rc},n,k}$ are frequency dependent as indicated by the subscript $n$, which is standard in multicarrier systems.

In the first sum of \eqref{equ:y_c1s}, $\alpha'_{\text{cc},n,1}$ refers to the direct desired signal depicted in Fig.\,\ref{fig:configuration}, i.e., the line of sight (LOS) path between the communication TX and RX. Meanwhile, in the second sum of \eqref{equ:y_c1s}, $\beta'_{\text{rc},n,1}$ refers to the direct cross interference from the radar TX to the communication RX. This is usually the strongest interference to the communication RX induced by spectrum sharing.

\emph{Assumption 1:} The propagation delay spread $\Delta\tau$, i.e., the difference between the smallest delay and largest delay, from the communication/radar TXs to the communication/radar RX is small with respect to (w.r.t.) the pulse duration $T$.

Assumption 1 is usually satisfied in a multicarrier system since each subcarrier is a narrowband system with a bandwidth $\Delta f\ll1/\Delta\tau$ \cite[Section 12.1]{Goldsmith2005}. In other words, Assumption 1 implies $\vert\tau_{\text{cc},k}-\tau_{\text{cc},1}\vert\ll T$, for $k>1$, and $\vert\widetilde{\tau}_{\text{rc},k}-\tau_{\text{cc},1}\vert\ll T$, $\forall k$, in \eqref{equ:y_c1s}.

After down-conversion, $y_{\text{c},n}(t)$ passes through a matched filter (MF) matched to the LOS communication waveform $q_\text{c}(t-\tau_{\text{cc},1})$ and is sampled at the symbol rate, which yields \cite[Section 5.1]{proakis2001digital}
\ben
\label{equ:y_c2s}
y_{\text{c},n}=\alpha_{\text{cc},n}d_n\sqrt{p_{\text{c},n}}+\beta_{\text{rc},n}\sqrt{p_{\text{r},n}}+w_{\text{c},n},
\een
where
\begin{align}
\alpha_{\text{cc},n}&=\int_{T}\sum_{k=1}^{K_{\text{cc}}}\alpha'_{\text{cc},n,k}q_\text{c}(t-\tau_{\text{cc},k})q_\text{c}^{\ast}(t-\tau_{\text{cc},1})dt,\\
\beta_{\text{rc},n}&=\int_{T}\sum_{k=1}^{K_{\text{rc}}}\beta'_{\text{rc},n,k}q_\text{r}(t-\widetilde{\tau}_{\text{rc},k})q_\text{c}^{\ast}(t-\tau_{\text{cc},1})dt,
\end{align}
and $w_{\text{c},n}$ is a zero-mean noise with
variance $\sigma_\text{c}^2$.

Now consider the radar received signal. Although the target is illuminated by both the radar TX and communication TX, we assume

\emph{Assumption 2:} The target echo due to the illumination from the communication source is negligible.

This assumption is valid because the communication source usually employs an omni-directional antenna for transmission, which leads to a much weaker target reflection compared with the target reflection due to the illumination from a directional radar TX.

Suppose there is a moving target located at range $R$ from the radar with a target radial velocity $v$. The round-trip delay between the radar and target is $\tau_{\text{rr}}=2R/c$, where $c$ is the speed of light. Then, the received signal at the radar RX on the $n$-th subcarrier can be written as \cite{SenNehorai2011}
\begin{align}
\label{equ:y_r1s}
y_{\text{r},n}(t)&=\bar{\alpha}\alpha'_{\text{rr},n}x_{\text{r},n}\big(\varepsilon(t-\tau_{\text{rr}})\big)+\sum_{k=1}^{K_{\text{rr}}}\beta'_{\text{rr},n,k}x_{\text{r},n}(t-\widetilde{\tau}_{\text{rr},k})\notag\\
&+\sum_{k=1}^{K_{\text{cr}}}\beta'_{\text{cr},n,k}x_{\text{c},n}(t-\widetilde{\tau}_{\text{cr},k})+w'_{\text{r},n}(t),
\end{align}
where $\bar{\alpha}$ is the radar cross-section (RCS), $\alpha'_{\text{rr},n}$ is a complex quantity representing the channel coefficient of the target path, $\varepsilon=1+\frac{2v}{c}$ is a scaling factor for the target Doppler shift, $\beta'_{\text{rr},n,k}$ denotes the complex scattering coefficient of the \mbox{$k$-th} clutter scatterer due to radar illumination with propagation delay $\widetilde{\tau}_{\text{rr},k}$, $\beta'_{\text{cr},n,k}$ and $\widetilde{\tau}_{\text{cr},k}$ are the scattering coefficient and, respectively, propagation delay associated with the $k$-th clutter scatterer due to the communication illumination, $K_{\text{rr}}$ and $K_{\text{cr}}$ are the total numbers of clutter scatterers observed at the radar RX due to the illumination of the radar TX and, respectively, communication TX, and $w'_{\text{r},n}(t)$ is the additive channel noise. Note that the direct-path interference from the communication TX to the radar RX is included as the first term of the second sum in \eqref{equ:y_r1s}, with $\widetilde{\tau}_{\text{cr},1}$ corresponding to the propagation delay between the communication TX and radar RX.

The radar signal $y_{\text{r},n}(t)$ is down-converted, Doppler compensated, filtered by a MF matched to the radar waveform $q_\text{r}(t-\tau_{\text{rr}})$, and sampled at the pulse rate. Like in the communication system, the propagation spread is assumed to be relatively small compared with the pulse duration $T$. The MF output can be written as \cite[Section 4.2]{richards2014fundamentals}:

\begin{align}
\label{equ:y_r2s}
y_{\text{r},n}=\alpha_{\text{rr},n}\sqrt{p_{\text{r},n}}+\beta_{\text{rr},n}\sqrt{p_{\text{r},n}}+\beta_{\text{cr},n}d_n\sqrt{p_{\text{c},n}}+w_{\text{r},n},
\end{align}
where
\begin{align}
\alpha_{\text{rr},n}=&\bar{\alpha}\int_{T}\alpha'_{\text{rr},n}q_\text{r}(t-\tau_{\text{rr}})q_\text{r}^{\ast}(t-\tau_{\text{rr}})dt,\\
\beta_{\text{rr},n}=&\int_{T}\sum_{k=1}^{K_{\text{rr}}}\beta'_{\text{rr},n,k}q_\text{r}(t-\widetilde{\tau}_{\text{rr},k})q_\text{r}^{\ast}(t-\tau_{\text{rr}})dt,\\
\beta_{\text{cr},n}=&\int_{T}\sum_{k=1}^{K_{\text{cr}}}\beta'_{\text{cr},n,k}q_\text{c}(t-\widetilde{\tau}_{\text{cr},k})q_\text{r}^{\ast}(t-\tau_{\text{rr}})dt,
\end{align}
and $w_{\text{r},n}$ is the output noise with zero mean and variance $\sigma_\text{r}^2$.

In this paper, the problem of interest is to jointly design the power allocation vectors $\pbf_\text{r}$ and $\pbf_\text{c}$ based on the radar-communication coexistence model in \eqref{equ:y_c2s} and \eqref{equ:y_r2s}.

\section{Proposed Approaches}
\label{sec:proposed_approach}
In this section, we propose two power allocation designs for the coexistence problem. The first one is a joint design, which considers the case when the radar and communication systems are fully cooperative, i.e., parameters of both systems are jointly designed to tackle the cross-interference induced by coexistence. The second one is a unilateral design, which is useful when the communication system is the primary user of the frequency band and the radar system wants to join and co-exist as a secondary user.
%

\subsection{Joint Design}
\label{subsec:jointdesign}
The figure of merit for the communication system is the achievable channel throughput, which is given by
\begin{gather}
C(\pbf_\text{r},\pbf_\text{c})=\sum_{n=1}^N\log_2\Big(1+\frac{\gamma_{\text{cc},n}p_{\text{c},n}}{\eta_{\text{rc},n}p_{\text{r},n}+1}\Big),
\end{gather}
where $\gamma_{\text{cc},n}=\frac{\mathbb{E}\{\vert\alpha_{\text{cc},n}\vert^2\}}{\sigma_\text{c}^2}$ and
$\eta_{\text{rc},n}=\frac{\mathbb{E}\{\vert\alpha_{\text{rc},n}\vert^2\}}{\sigma_\text{c}^2}$ denote the \emph{normalized
  signal-to-noise ratio} (SNR) and \emph{normalized
  interference-to-noise ratio} (INR) at the communication
receiver, which are effectively the SNR or INR per unit transmission power. $\mathbb{E}\{\cdot\}$
represents the statistical expectation. For the radar system, the figure of merit is the SINR
\begin{gather}
\text{SINR}(\pbf_\text{r},\pbf_\text{c})=\sum_{n=1}^N\frac{\gamma_{\text{rr},n}p_{\text{r},n}}{\eta_{\text{rr},n}p_{\text{r},n}+\eta_{\text{cr},n}p_{\text{c},n}+1},
\end{gather}
where $\gamma_{\text{rr},n}=\frac{\mathbb{E}\{\vert\alpha_{\text{rr},n}\vert^2\}}{\sigma_\text{r}^2}$, $\eta_{\text{rr},n}=\frac{\mathbb{E}\{\vert\beta_{\text{rr},n}\vert^2\}}{\sigma_\text{r}^2}$, and
$\eta_{\text{cr},n}=\frac{\mathbb{E}\{\vert\beta_{\text{cr},n}\vert^2\}}{\sigma_\text{r}^2}$ are the normalized SNR, \emph{clutter-to-noise ratio} (\mbox{CNR}), and INR, respectively. The joint power allocation problem is formulated as maximizing the radar SINR under throughput and power constraints:
\begin{subequations}
\label{equ:P1}
\begin{gather}
\label{equ:sharing_abj}
\max\limits_{\pbf_\text{r},\pbf_\text{c}}~~\text{SINR}(\pbf_\text{r},\pbf_\text{c}),
\\
\label{equ:sharing_c1}
\text{s.t.}~\sum_{n=1}^Np_{\text{r},n}\leq P_\text{r},~\sum_{n=1}^N p_{\text{c},n}\leq P_\text{c},
\\
\label{equ:sharing_c2}
0\leq p_{\text{r},n}\leq\xi_\text{r},~0\leq p_{\text{c},n}\leq \xi_\text{c},~\forall~n,
\\
\label{equ:sharing_c3}
C(\pbf_\text{r},\pbf_\text{c})\geq \kappa,
\end{gather}
\end{subequations}
where \eqref{equ:sharing_c1} represents the total transmission power constraint for each system, \eqref{equ:sharing_c2} denotes subchannel peak power constraints, and \eqref{equ:sharing_c3} is a communication throughput constraint.


The joint design problem \eqref{equ:P1} is nonconvex since the objective function and the constraint \eqref{equ:sharing_c3} are both nonconvex. The above problem may be solved by employing an alternating optimization procedure \cite{AubryMaioTSP18}. The idea is to iteratively solve \eqref{equ:P1} w.r.t. $\pbf_r$ while keeping $\pbf_c$ fixed, and vice versa, until convergence is reached. However, this alternating maximization method is computationally intensive and does not guarantee convergence. This is particularly so for the considered cluttered environment, where the clutter term in the SINR depends on the power allocation variable $\pbf_\text{r}$, which makes the optimization problem significantly more challenging even with fixed $\pbf_\text{c}$. To address these challenges, we consider a different approach that is described next.

Specifically, let us define $\etabf_{\text{c},n}=[\eta_{\text{rc},n},\ 0]^T$, $\etabf_{\text{r},n}=[\eta_{\text{rr},n},\ \eta_{\text{cr},n}]^T$, $\gammabf_{\text{r},n}=[\gamma_{\text{rr},n},\ 0]^T$, $\gammabf_{\text{c},n}=[0,\ \gamma_{\text{cc},n}]^T$, and $\Pbf=[\pbf_{\text{r}}^T;\ \pbf_{\text{c}}^T]$ is a $2\times N$ matrix. Then, \eqref{equ:P1} can be rewritten as
\begin{subequations}
\label{equ:NonP1}
\begin{gather}
\label{equ:sharingNon_abj}
\max\limits_{\Pbf}~~\sum_{n=1}^N\frac{\gammabf_{\text{r},n}^T\Pbf\sbf_n}{\etabf_{\text{r},n}^T\Pbf\sbf_n+1},
\\
\label{equ:sharingNon_c1}
\text{s.t.}~\eqref{equ:sharing_c1},~\eqref{equ:sharing_c2},
\\
\label{equ:sharingNon_c3}
\sum_{n=1}^N\log_2\Big(1+\frac{\gammabf_{\text{c},n}^T\Pbf\sbf_n}{\etabf_{\text{c},n}^T\Pbf\sbf_n+1}\Big)\geq \kappa,
\end{gather}
\end{subequations}
where $\sbf_n$ is a $N\times1$ selection vector given by
\ben
\sbf_n(i)=
\begin{cases}
  1,~i=n,\\
  0,~\text{otherwise}.
\end{cases}
\een
Note that \eqref{equ:NonP1} is a fractional programming (FP) with the objective function being a sum of multiple ratios.

\emph{Remark 1}: The conventional alternating optimization approach usually decomposes the original nonconvex problem \eqref{equ:P1} into two subproblems in $\pbf_r$ and $\pbf_c$, respectively \cite{LiPetropuluTSP16,LiPetropulu2017,ZhengLopsWangTSP2018,AubryMaioTSP18,RihanHuang18,ChengLiaoHe19}. Although the subproblems are simpler than the original problem, they are still nonconvex and require convex relaxation techniques to solve. Specifically, when $\pbf_c$ is fixed, the subproblem in $\pbf_r$ has a similar form as \eqref{equ:NonP1}, which is multiple-ratio FP problem. On the other hand, when fixing $\pbf_r$, the resulting subproblem is also nonconvex. The alternating approach needs to solve both nonconvex subproblems multiple times till convergence or a fixed number of iterations is completed. To bypass the alternating procedure, we combine the design variables into a single stacked variable and transfer the original problem into a simplified form. A direct benefit is computational saving since we need to solve the multiple-ratio FP problem only once. Simulation results show that the complexity of the proposed algorithm is considerably lower than that of the alternating procedure.

The multiple-ratio FP problem \eqref{equ:NonP1} is nonconvex since the objective function is a sum of ratios, which is nonconvex, and the throughput constraint \eqref{equ:sharingNon_c3} imposes a nonconvex feasible set. To solve \eqref{equ:NonP1}, we can reformulate the objective function and employ an inner iteration based on convex relaxation for the throughput constraint. First, for the objective function, a quadratic transform can be used \cite{ShenYu2018}. This approach introduces a set of slack variables $\lambdabf=[\lambda_1,\cdots,\lambda_N]^T$ to deal with the nonconvexity. Specifically, problem \eqref{equ:NonP1} is equivalent to
\begin{subequations}
\label{equ:NonP2}
\begin{gather}
\label{equ:optNon_sr_new}
\max\limits_{\Pbf,\lambdabf}~~F(\lambdabf,\Pbf),
\\
\text{s.t.}~\eqref{equ:sharing_c1},~\eqref{equ:sharing_c2},~\eqref{equ:sharingNon_c3},
\end{gather}
\end{subequations}
where
\begin{align}
F(\lambdabf,\Pbf)=\sum_{n=1}^N\Big(2\lambda_n\sqrt{\gammabf_{\text{r},n}^T\Pbf\sbf_n}-\lambda_n^2\big(\etabf_{\text{r},n}^T\Pbf\sbf_n+1\big)\Big).
\end{align}
Let $\lambda_n^{(\ell-1)}$ and $\tilde{\Pbf}^{(\ell-1)}$ denote the solutions obtained from the $(\ell-1)$-st iteration. Then, $\lambda_n^{(\ell)}$ can be updated by solving the following problem:
\ben
\max\limits_{\lambdabf}~~F(\lambdabf,\tilde{\Pbf}^{(\ell-1)}),
\een
which has a closed-form solution:
\ben\label{equ:lambdacomp}
\lambda_n^{(\ell)}=\frac{\sqrt{\gammabf_{\text{r},n}^T\tilde{\Pbf}^{(\ell-1)}\sbf_n}}{\etabf_{\text{r},n}^T\tilde{\Pbf}^{(\ell-1)}\sbf_n+1}.
\een
In turn, $\tilde{\Pbf}^{(\ell)}$ can be obtained by solving
\begin{subequations}
\label{equ:NonP3}
\begin{gather}
\label{equ:opt_sr_Nonp3}
\max\limits_{\Pbf}~~F(\lambdabf^{(\ell)},\Pbf),
\\
\text{s.t.}~\eqref{equ:sharing_c1},~\eqref{equ:sharing_c2},~\eqref{equ:sharingNon_c3},
\end{gather}
\end{subequations}
Note that the above problem is nonconvex since \eqref{equ:sharingNon_c3} imposes a nonconvex set. We can use a SCP process to relax constraint \eqref{equ:sharingNon_c3} by converting it into a convex set along with an inner iteration to solve \eqref{equ:NonP3}. Specifically, \eqref{equ:sharingNon_c3} can be relaxed into a linear form as
\ben\label{equ:consrate}
\sum_{n=1}^{N}\log_2\big(\gammabf_{\text{c},n}^T\Pbf\sbf_n+\etabf_{\text{c},n}^T\Pbf\sbf_n+1\big)-G(\Pbf,\hat{\Pbf}^{(\ell_\text{s}-1)})\geq\kappa,
\een
where $\hat{\Pbf}^{(\ell_\text{s}-1)}$ is the power vector from the $(\ell_\text{s}-1)$-st inner SCP iteration and
\begin{align}\label{equ:consscpinner}
G(\Pbf,\hat{\Pbf}^{(\ell_\text{s}-1)})&\triangleq\log_2(\etabf_{\text{c},n}^T\hat{\Pbf}^{(\ell_\text{s}-1)}\sbf_n+1)\notag\\
&+\frac{\etabf_{\text{c},n}^T(\Pbf-\hat{\Pbf}^{(\ell_\text{s}-1)})\sbf_n}{\ln2(\etabf_{\text{c},n}^T\hat{\Pbf}^{(\ell_\text{s}-1)}\sbf_n+1)}.
\end{align}
Thus, during the $\ell_\text{s}$-th inner SCP iteration, the following convex optimization problem is solved to obtain $\hat{\Pbf}^{(\ell_\text{s})}$:
\begin{subequations}
\label{equ:NonP4}
\begin{gather}
\label{equ:opt_sr_Nonp4}
\max\limits_{\Pbf}~~F(\lambdabf^{(\ell)},\Pbf),
\\
\text{s.t.}~\eqref{equ:sharing_c1},~\eqref{equ:sharing_c2},~\eqref{equ:consrate}.
\end{gather}
\end{subequations}
After convergence, $\tilde{\Pbf}^{(\ell)}=\hat{\Pbf}^{(\ell_\text{s})}$ is used in \eqref{equ:lambdacomp} to compute $\lambda_n$ for the next quadratic transform iteration. Our proposed solution to the joint design problem is summarized in $\textbf{Algorithm~\ref{alg:Joint}}$.

The computational complexity of $\textbf{Algorithm~\ref{alg:Joint}}$ depends on the number of the quadratic transform iterations $L$ as well as the number of the SCP iterations $L_\text{s}$. Simulations show that the required number of the inner or outer iteration is relatively small. In addition, the convex problem \eqref{equ:NonP4} inside the iteration has a complexity of $\Ocal(N^{3.5})$ when an interior-point method is used \cite{Boyd2004}. Thus, the overall complexity of the proposed solution is $\Ocal(LL_\text{s}N^{3.5})$.

\begin{algorithm}[t]
\caption{Proposed Joint Design}
\begin{algorithmic}
\label{alg:Joint}
\STATE \textbf{Input:} Channel SNRs $\gamma_{\text{rr},n}$ and $\gamma_{\text{cc},n}$, channel INRs $\eta_{\text{rc},n}$ and $\eta_{\text{cr},n}$, CNR $\eta_{\text{rr},n}$, total powers $P_\text{r}$ and $P_\text{c}$, peak power constraints $\xi_\text{r}$ and $\xi_\text{c}$, throughput constraint $\kappa$, and tolerance $\epsilon$.
\STATE \textbf{Output:} Radar and communication powers $\Pbf$.\\
\STATE  \textbf{Initialization:} Initialize $\tilde{\Pbf}^{(0)}$ and set iteration index $\ell=0$.\\
\REPEAT
\STATE
\begin{enumerate}
  \item Set $\ell=\ell+1$.
  \item Solve problem \eqref{equ:lambdacomp} to obtain $\lambda_n^{(\ell)}$.
  \item Initialization: $\ell_\text{s}=0$ and $\hat{\Pbf}^{(\ell_\text{s})}=\tilde{\Pbf}^{(\ell-1)}$.
        \STATE \textbf{repeat}
        \begin{enumerate}
         \item Set $\ell_\text{s}=\ell_\text{s}+1$.
          \item Solve problem \eqref{equ:NonP4} with fixed $\hat{\Pbf}^{(\ell_\text{s}-1)}$ and $\lambda_n^{(\ell)}$ to obtain $\hat{\Pbf}^{(\ell_\text{s})}$.
        \end{enumerate}
        \STATE \textbf{until} convergence.
  \item Update $\tilde{\Pbf}^{(\ell)}=\hat{\Pbf}^{(\ell_\text{s})}$.
\end{enumerate}
\UNTIL convergence.
\RETURN $\Pbf=\tilde{\Pbf}^{(\ell)}$.
\end{algorithmic}
\end{algorithm}
\subsection{Unilateral Design}
\label{subsec:UniDesign}
The above joint design requires mutual cooperation of both radar and communication systems. However, in some scenarios, the communication system may be the primary and pre-existing user of the frequency band, while the radar occasionally joins and co-exists with the primary user. Thus, we consider a second spectrum sharing framework based on a unilateral design from the radar perspective that optimizes the radar transmission power $\pbf_{\text{r}}$ when the communication transmission power is known and fixed \cite{ShiSellathurai2018}.

Specifically, suppose the communication system pre-exists and employs a waterfilling approach to allocate subchannel power before the radar enters the channel:
\begin{subequations}
\label{equ:P_u}
\begin{gather}
\label{equ:sharing_abj_u}
\tilde{\pbf}_\text{c}=\arg~\max\limits_{\pbf_\text{c}}~~\sum_{n=1}^N\log_2\Big(1+\gamma_{\text{cc},n}p_{\text{c},n}\Big),
\\
\label{equ:sharing_c1_u}
\text{s.t.}~\sum_{n=1}^N p_{\text{c},n}\leq P_\text{c},~0\leq p_{\text{c},n}\leq \xi_\text{c},~\forall~n,
\end{gather}
\end{subequations}
which is convex and can be solved by waterfilling.

When the radar needs to access the channel, it acquires the knowledge of communication power allocation and uses a strategy to maximize its SINR, subject to a minimum communication throughput constraint and power constraints:
\begin{subequations}
\label{equ:P4}
\begin{gather}
\label{equ:opt_ad-scp}
\max\limits_{\pbf_\text{r}}~~\sum_{n=1}^N\frac{\gamma_{\text{rr},n}p_{\text{r},n}}{\eta_{\text{rr},n}p_{\text{r},n}+\eta_{\text{cr},n}\tilde{p}_{\text{c},n}+1},
\\
\label{equ:c1_ad-scp}
\text{s.t.}~0\leq p_{\text{r},n}\leq \xi_\text{r},~\forall~n,~\sum_{n=1}^Np_{\text{r},n}\leq P_\text{r},
\\
\label{equ:c2_ad-scp}
\sum_{n=1}^N\log_2\Big(1+\frac{\gamma_{\text{cc},n}\tilde{p}_{\text{c},n}}{\eta_{\text{rc},n}p_{\text{r},n}+1}\Big)\geq \kappa.
\end{gather}
\end{subequations}
The objective function can be rewritten as
\ben
\sum_{n=1}^{N}\frac{\gamma_{\text{rr},n}}{\eta_{\text{rr},n}}-\sum_{n=1}^{N}\frac{\gamma_{\text{rr},n}(\eta_{\text{cr},n}\tilde{p}_{\text{c},n}+1)}{\eta_{\text{rr},n}^2p_{\text{r},n}+\eta_{\text{rr},n}(\eta_{\text{cr},n}\tilde{p}_{\text{c},n}+1)}.
\een
Thus, problem \eqref{equ:P4} is equivalent to
\begin{subequations}
\label{equ:P6}
\begin{gather}
\label{equ:opt_sr_p6}
\min\limits_{\pbf_\text{r}}~~\sum_{n=1}^{N}\frac{\gamma_{\text{rr},n}(\eta_{\text{cr},n}\tilde{p}_{\text{c},n}+1)}{\eta_{\text{rr},n}^2p_{\text{r},n}+\eta_{\text{rr},n}(\eta_{\text{cr},n}\tilde{p}_{\text{c},n}+1)},
\\
\text{s.t.}~\eqref{equ:c1_ad-scp},~\eqref{equ:c2_ad-scp},
\end{gather}
\end{subequations}
Note that while the objective \eqref{equ:opt_sr_p6} is convex, the above problem is nonconvex since \eqref{equ:c2_ad-scp} is a nonconvex set. We can use the first-order Taylor expansion to convert the nonconvex constraint into a convex one and solve the relaxed problem using an SCP process. Specifically, rewrite the left side of \eqref{equ:c2_ad-scp} as
\ben
\sum_{n=1}^Nf(p_{\text{r},n}),
\een
where
\ben\label{equ:dc_obj}
f(p_{\text{r},n})\triangleq F_1(p_{\text{r},n}\vert \tilde{p}_{\text{c},n})-\log_2\big(\eta_{\text{rc},n}p_{\text{r},n}+1\big),
\een
and
\ben
F_1(p_{\text{r},n}\vert \tilde{p}_{\text{c},n})=\log_2\big(\eta_{\text{rc},n}p_{\text{r},n}+1+\gamma_{\text{cc},n}\tilde{p}_{\text{c},n}\big).
\een
It can be shown that $f(p_{\text{r},n})$ is nonconvex w.r.t. $p_{\text{r},n}$ since it is a difference of two concave functions. The second concave function in \eqref{equ:dc_obj} can be upper bounded by a first-order Taylor expansion at $\tilde{p}_{\text{r},n}^{(\ell_\text{r}-1)}$:
\ben\label{equ:F2approx}
\begin{split}
&\log_2\big(\eta_{\text{rc},n}p_{\text{r},n}+1\big)\leq F_2(p_{\text{r},n}\vert\tilde{p}_{\text{r},n}^{(\ell_\text{r}-1)})\\&\triangleq\log_2\big(\eta_{\text{rc},n}\tilde{p}_{\text{r},n}^{(\ell_\text{r}-1)}+1\big)
+\frac{\eta_{\text{rc},n}(p_{\text{r},n}-\tilde{p}_{\text{r},n}^{(\ell_\text{r}-1)})}{\ln 2(\eta_{\text{rc},n}\tilde{p}_{\text{r},n}^{(\ell_\text{r}-1)}+1)},
\end{split}
\een
where $\tilde{p}_{\text{r},n}^{(\ell_\text{r}-1)}$ is the radar power from the $(\ell_\text{r}-1)$-st inner iteration. Clearly, the bound is tight at $p_{\text{r},n}=\tilde{p}_{\text{r},n}^{(\ell_\text{r}-1)}$:
\ben\label{equ:tightbound}
\log_2\big(\eta_{\text{rc},n}\tilde{p}_{\text{r},n}^{(\ell_\text{r}-1)}+1\big)= F_2(\tilde{p}_{\text{r},n}^{(\ell_\text{r}-1)}\vert\tilde{p}_{\text{r},n}^{(\ell_\text{r}-1)}).
\een
Substituting \eqref{equ:F2approx} back into \eqref{equ:dc_obj} gives the lower bound of $f(p_{\text{r},n})$: $\tilde{f}(p_{\text{r},n})=F_1(p_{\text{r},n}\vert \tilde{p}_{\text{c},n})-F_2(p_{\text{r},n}\vert\tilde{p}_{\text{r},n}^{(\ell_\text{r}-1)})$. We can see that $\tilde{f}(p_{\text{r},n})$ is now an affine function of $p_{\text{r},n}$ and constraint \eqref{equ:c2_ad-scp} becomes
\ben\label{equ:relaxc}
\sum_{n=1}^N\tilde{f}(p_{\text{r},n})\geq \kappa,
\een
which is a convex set. Thus, during the $\ell_\text{r}$-th inner SCP iteration, the following convex problem is solved for $\tilde{p}_{\text{r},n}^{(\ell_\text{r})}$ until convergence:
\begin{subequations}
\label{equ:P7}
\begin{gather}
\label{equ:opt_sr_p7}
\min\limits_{\pbf_\text{r}}~~\sum_{n=1}^{N}\frac{\gamma_{\text{rr},n}(\eta_{\text{cr},n}\widetilde{p}_{\text{c},n}+1)}{\eta_{\text{rr},n}^2p_{\text{r},n}+\eta_{\text{rr},n}(\eta_{\text{cr},n}\widetilde{p}_{\text{c},n}+1)},
\\
\text{s.t.}~\eqref{equ:c1_ad-scp},~\eqref{equ:relaxc}.
\end{gather}
\end{subequations}
The proposed solution is summarized in $\textbf{Algorithm~\ref{alg:Unilateral}}$.

Similar to $\textbf{Algorithm~\ref{alg:Joint}}$, the complexity of $\textbf{Algorithm~\ref{alg:Unilateral}}$  depends on the number of the SCP iterations $L_\text{r}$ required for convergence, and the overall computational complexity is $\Ocal(L_\text{r}N^{3.5})$.
\begin{algorithm}[t]
\caption{Proposed Unilateral Design}
\begin{algorithmic}
\label{alg:Unilateral}
\STATE \textbf{Input:} $\gamma_{\text{rr},n}$, $\gamma_{\text{cc},n}$, $\eta_{\text{rc},n}$, $\eta_{\text{cr},n}$, $\eta_{\text{rr},n}$, $P_\text{r}$, $\widetilde{\pbf}_\text{c}$, $\xi_\text{r}$, and $\kappa$ (same as in $\textbf{Algorithm~\ref{alg:Joint}}$).
\STATE \textbf{Output:} Radar powers $\pbf_\text{r}$.\\
\STATE  \textbf{Initialization:} Initialize $\tilde{p}_{\text{r},n}^{(0)}$ and set iteration index $\ell_\text{r}=0$.\\

\REPEAT
\STATE
\begin{enumerate}
  \item Set $\ell_\text{r}=\ell_\text{r}+1$.
  \item Solve problem \eqref{equ:P7} to obtain $\tilde{p}_{\text{r},n}^{(\ell_\text{r})}$.
\end{enumerate}
\UNTIL convergence.
\RETURN $p_{\text{r},n}=\tilde{p}_{\text{r},n}^{(\ell_\text{r})}$.
\end{algorithmic}
\end{algorithm}
\subsection{Feasibility and Initialization Analysis}
\label{subsec:Feasibility}
For the joint design problem \eqref{equ:P1}, its feasibility depends on if the maximum achievable throughput (e.g., $C_{\text{max}}$) under the power constraints is no less than the minimum throughput constraint $\kappa$, that is, $C_{\text{max}}\geq\kappa$. Clearly, $C_{\text{max}}$ is achieved when the radar is absent while the communication system uses all subcarriers to maximize its throughput, which is the same as problem \eqref{equ:P_u}. In other words, problem \eqref{equ:P1} is feasible if the following condition is statisfied:
\ben
\label{equ:feasibility}
\sum_{n=1}^N\log_2\Big(1+\gamma_{\text{cc},n}\tilde{p}_{\text{c},n}\Big)\geq\kappa.
\een
It is easy to show that \eqref{equ:feasibility} provides also the feasible condition for the unilateral design.

Note that the proposed solutions for the joint and unilateral design requires initial values of $\pbf_c$ and, respectively, $\pbf_\text{r}$. A simple way of initialization is to consider only the power constraints \eqref{equ:sharing_c1} and \eqref{equ:sharing_c2}. A better way that also takes into account the throughput constraint $\eqref{equ:sharing_c3}$ is a greedy search (GS) method, i.e., the communication system uses its best subcarriers to maintain the throughput constraint, while the radar employs the remaining subcarriers to maximize its SINR. The detailed steps of the GS method is summarized in $\textbf{Algorithm~\ref{alg:GS}}$.

\begin{algorithm}[t]
\caption{Greedy Search}
\begin{algorithmic}
\label{alg:GS}
\STATE \textbf{Input:} $\gamma_{\text{rr},n}$, $\gamma_{\text{cc},n}$, $\eta_{\text{rc},n}$, $\eta_{\text{cr},n}$, $\eta_{\text{rr},n}$, $P_\text{r}$, $P_\text{c}$, $\xi_\text{r}$, $\xi_\text{c}$, and $\kappa$.
\STATE \textbf{Output:} Radar and communication powers $\pbf_\text{r}$ and $\pbf_\text{c}$.\\
\begin{enumerate}
\item Define a binary selection vector $\ubf=[u_1,\cdots,u_N]^T$ with $u_n\in\{0,1\}$: $u_n=1$ indicates the communication system uses the $n$-th subcarrier; otherwise, the radar uses it.
\item Sort the normalized communication channel SNRs $\gamma_{cc,n}$ in a descending order. The SNRs after sorting are denoted by $\gamma_{\text{cc},n,m}$, where the subscripts $(n,m)$ indicate the indices of a subcarrier before and after sorting.
\item Set $\iota=0$, which denotes the partial sum of the communication throughput, $\ubf=0$, and $m=1$.
\REPEAT
\STATE
\begin{enumerate}
\item $u_n=1$ and $m=m+1$.
\item Solve the following convex problem and denote the solution by $\hat{\pbf}_\text{c}$:
\begin{subequations}
\label{equ:P_gs}
\begin{gather}
\label{equ:sharing_abj_gs}
\max\limits_{\pbf_\text{c}}~\sum_{n=1}^N\log_2\Big(1+u_n\gamma_{\text{cc},n,m}p_{\text{c},n}\Big),
\\
\label{equ:sharing_c1_gs}
\text{s.t.}~\sum_{n=1}^N u_np_{\text{c},n}\leq P_\text{c},\\~0\leq u_np_{\text{c},n}\leq \xi_\text{c},~\forall~n,
\end{gather}
\end{subequations}\\
\item Compute $\iota=\sum_{n=1}^N\log_2\big(1+u_n\gamma_{\text{cc},n,m}\hat{p}_{\text{c},n}\big)$.
\end{enumerate}
\UNTIL $\iota\geq\kappa$.
\item Compute the radar power $\hat{\pbf}_\text{r}$ by solving:
    \begin{subequations}
\label{equ:Pgs}
\begin{gather}
\label{equ:opt_sr_pgs}
\min\limits_{\pbf_\text{r}}~\sum_{n=1}^{N}\frac{\gamma_{\text{rr},n}(\eta_{\text{cr},n}u_n\hat{p}_{\text{c},n}+1)}{\eta_{\text{rr},n}^2(1-u_n)p_{\text{r},n}+\eta_{\text{rr},n}(\eta_{\text{cr},n}u_n\hat{p}_{\text{c},n}+1)},
\\
\text{s.t.}~\sum_{n=1}^N (1-u_n)p_{\text{r},n}\leq P_\text{r},\\~0\leq (1-u_n)p_{\text{r},n}\leq \xi_\text{r},~\forall~n,
\end{gather}
\end{subequations}
\end{enumerate}
\RETURN $\pbf_\text{c}=\hat{\pbf}_\text{c}\odot\ubf$ and $\pbf_\text{r}=(\mathbf{1}_{1\times N}-\ubf)\odot\hat{\pbf}_\text{r}$.
\end{algorithmic}
\end{algorithm}
\section{Numerical Simulations}
\label{sec:simulationresults}
In this section, numerical results are presented to illustrate the performance of different methods for spectrum sharing between multicarrier radar and communication systems. Specifically, we compare the proposed \textbf{joint design} in Section \ref{subsec:jointdesign} and \textbf{unilateral design} in Section \ref{subsec:UniDesign} with the heuristic \textbf{greedy search} method. In addition, we include the optimum radar output SINR, under the condition when the communication system is absent (denoted as \textbf{comm absent}), as an upper bound.

Unless stated otherwise, the number of subcarriers $N=16$, the convergence tolerance is $0.01$, the noise variance $\sigma^2_\text{r}=\sigma_\text{c}^2=1$, and the communication throughput constraint is $\kappa=2.5$ bits/s/Hz in most examples. The subcarrier channel coefficients
$\alpha_{\text{rr},n}$, $\alpha_{\text{cc},n}$, $\beta_{\text{rc},n}$, $\beta_{\text{rr},n}$, and $\beta_{\text{cr},n}$ are generated with Gaussian distribution
$\Ccal\Ncal(0,\sigma_{\text{rr}}^2)$, $\Ccal\Ncal(0,\sigma_{\text{cc}}^2)$,
$\Ccal\Ncal(0,\sigma_{\text{rc}}^2)$, $\Ccal\Ncal(0,\sigma^2)$ and $\Ccal\Ncal(0,\sigma_{\text{cr}}^2)$,
respectively. The strength of the desired signal for both systems,
indicated by $\sigma_{\text{rr}}^2$ and $\sigma_{\text{cc}}^2$, are normalized as
$\sigma_{\text{rr}}^2=\sigma_{\text{cc}}^2=1$. The clutter strength $\sigma^2=0.05$. In the sequel, we consider two coexistence scenarios characterized by the strength of the cross interference:
\begin{itemize}
\item Case 1 (weak cross interference): $\sigma_{\text{rc}}^2=\sigma_{\text{cr}}^2=0.01$.
\item Case 2 (strong cross interference): $\sigma_{\text{rc}}^2=\sigma_{\text{cr}}^2=0.1$.
\end{itemize}
In the simulation, 50 trials of channel realization are utilized to obtain the average performance.

\begin{figure}
\centering
\includegraphics[width=3.1in]{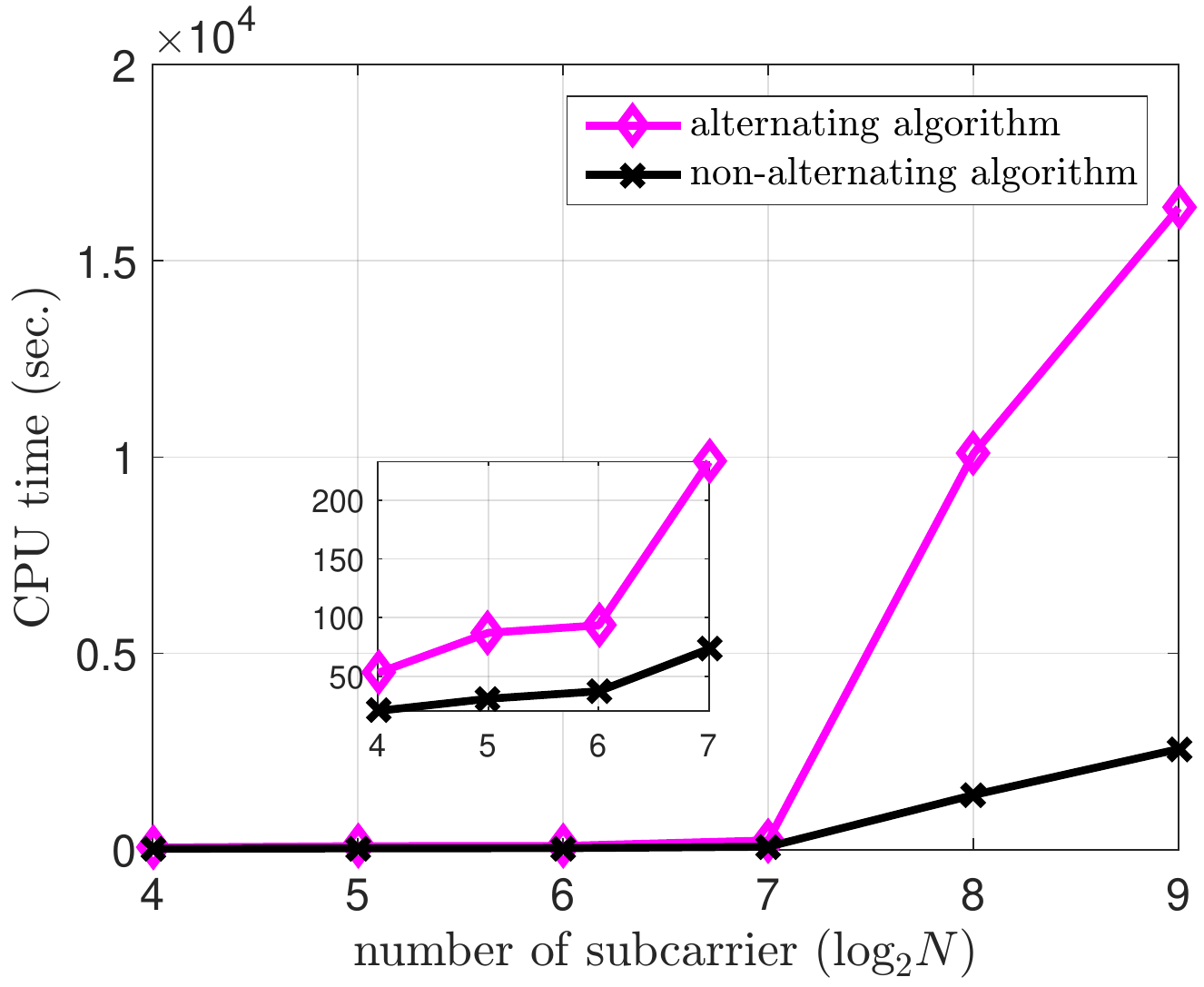}
\caption{Computer simulation time versus the number of subcarrier for the conventional alternating algorithm and the proposed non-alternating algorithm.}
\label{fig:CPUtime}
\end{figure}

First, we consider the computational complexity of the conventional alternating optimization approach, which decompose the original problem into two subproblems in $\pbf_\text{r}$ and $\pbf_\text{c}$, and the proposed non-alternating method as discussed in Section \ref{subsec:jointdesign}. Fig.~\ref{fig:CPUtime} shows the CPU time measured by Matlab versus the total number of subcarriers $N$ for Case 1, where $P_\text{r}=P_\text{c}=600$ and $\kappa=1.5$. It can be seen that the complexity of both methods grows as the number of subcarriers increase. However, the alternating algorithm is seen to take a longer time to converge for all cases considered. In particular, the alternating algorithm is around 8 times slower than the proposed non-alternating method at $N=512$.

\begin{figure}
\centering
\bt{cc}
\includegraphics[width=3in]{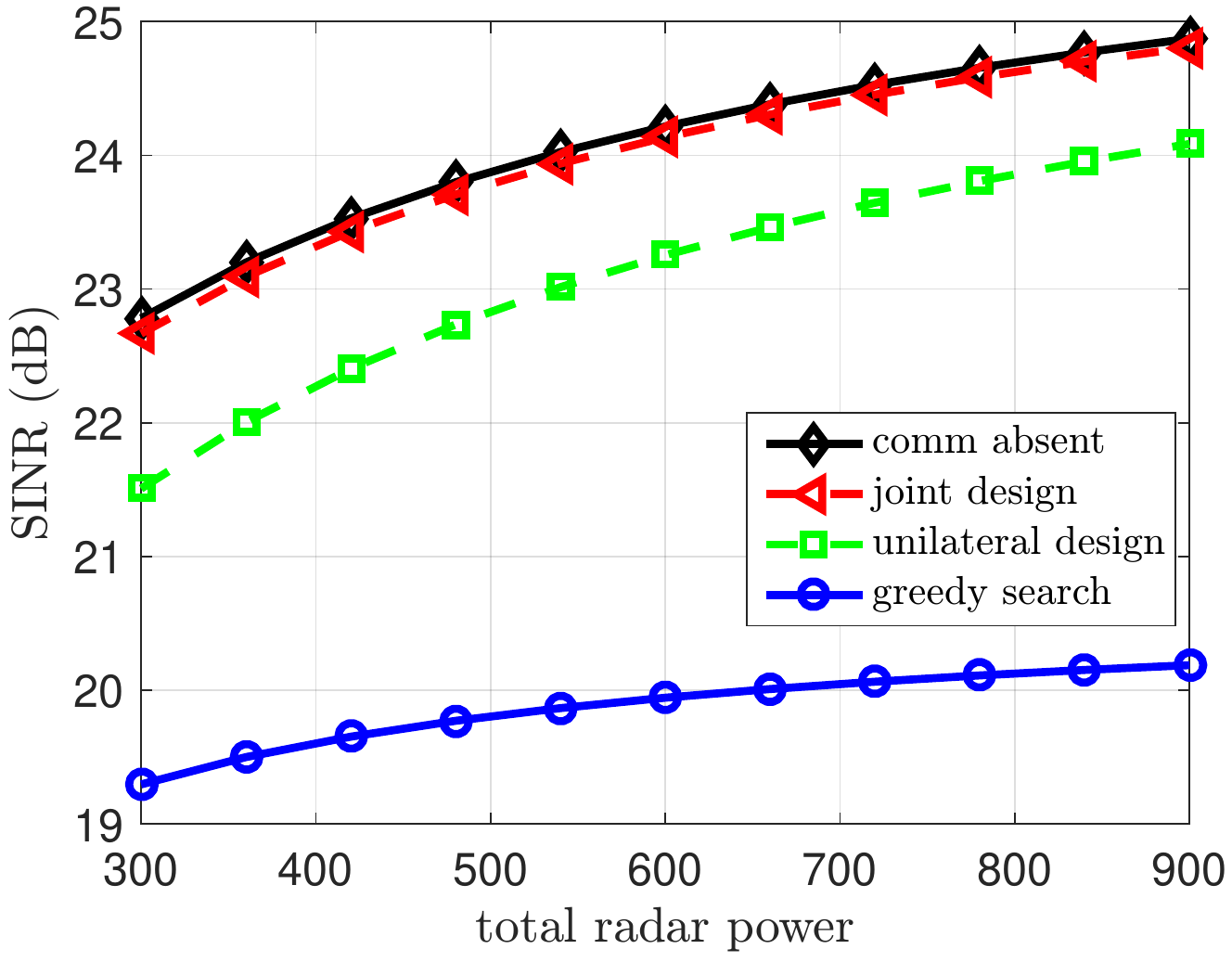}\\ 
(a)\\
\includegraphics[width=3in]{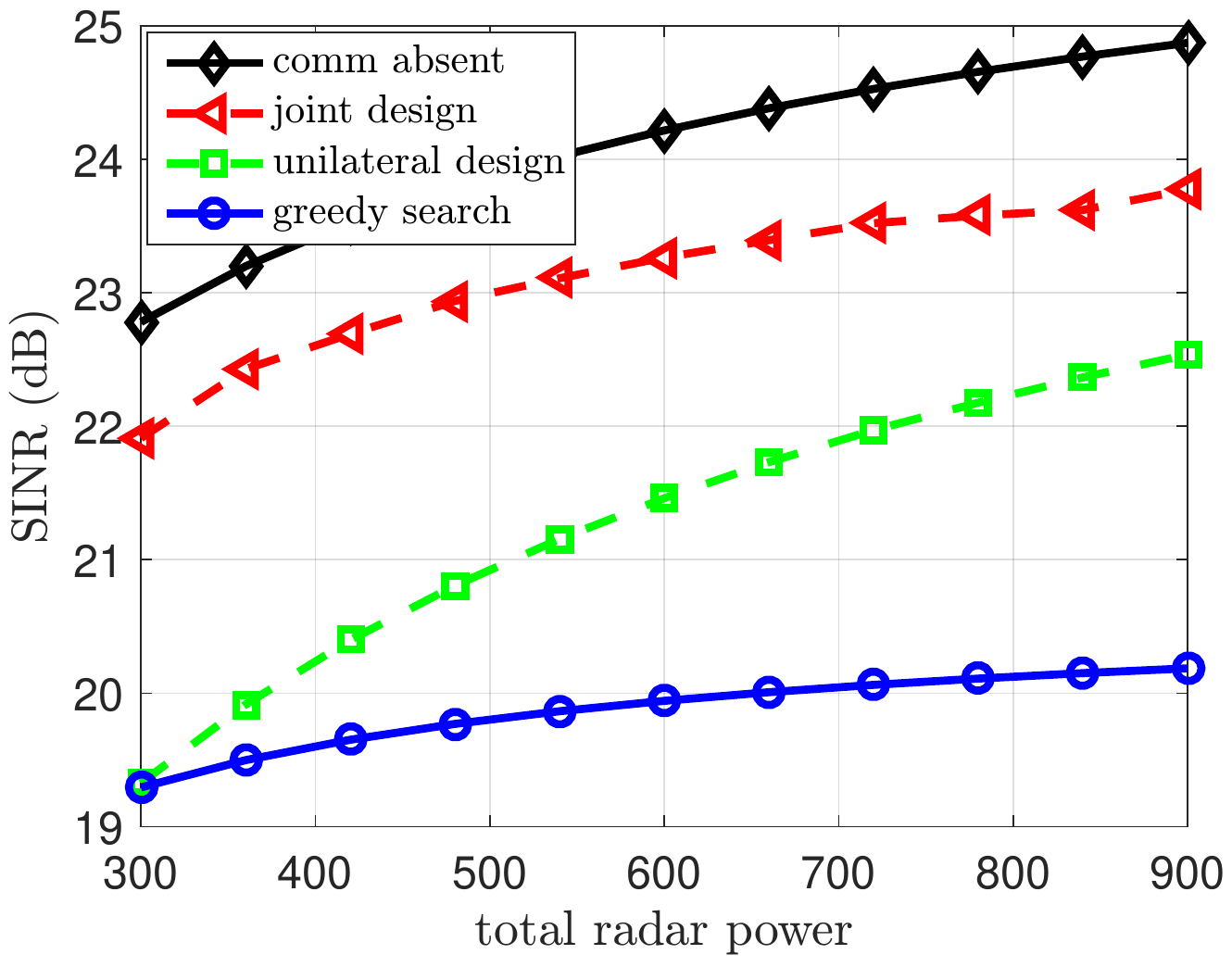}
\\
(b)
\et
\caption{Output SINR versus the total radar transmission power. (a) Weak cross interference (Case 1); (b) strong cross interference (Case 2).}
\label{fig:subfigureSINR}
\end{figure}

Fig.~\ref{fig:subfigureSINR} shows the output radar SINR versus the total radar transmission power when $P_{\text{c}}=600$ and $\kappa=2.5$ for Case 1 and Case 2, respectively. It can be seen from Fig.~\ref{fig:subfigureSINR} (a) that with weak cross interference (Case 1), the output SINR of the joint design is very close to that of the comm absent scenario since the weak cross interference creates limited impact from one system to the other. On the other hand, there is a notable performance loss for the radar unilateral design due to the fixed communication power allocation. When the cross interference gets stronger (Case 2), as indicated in Fig.~\ref{fig:subfigureSINR} (b), both the joint design and unilateral design degrade, although the unilateral design experiences a more severe performance loss. In both Case 1 and Case 2, the joint design and unilateral design, which involve subcarrier sharing between the radar and communication systems, outperform the GS method, which is a subcarrier-allocation based method. As the total radar transmission power increases, the output SINR of all considered scenarios increases.

\begin{figure}
\centering
\bt{cc}
\includegraphics[width=3in]{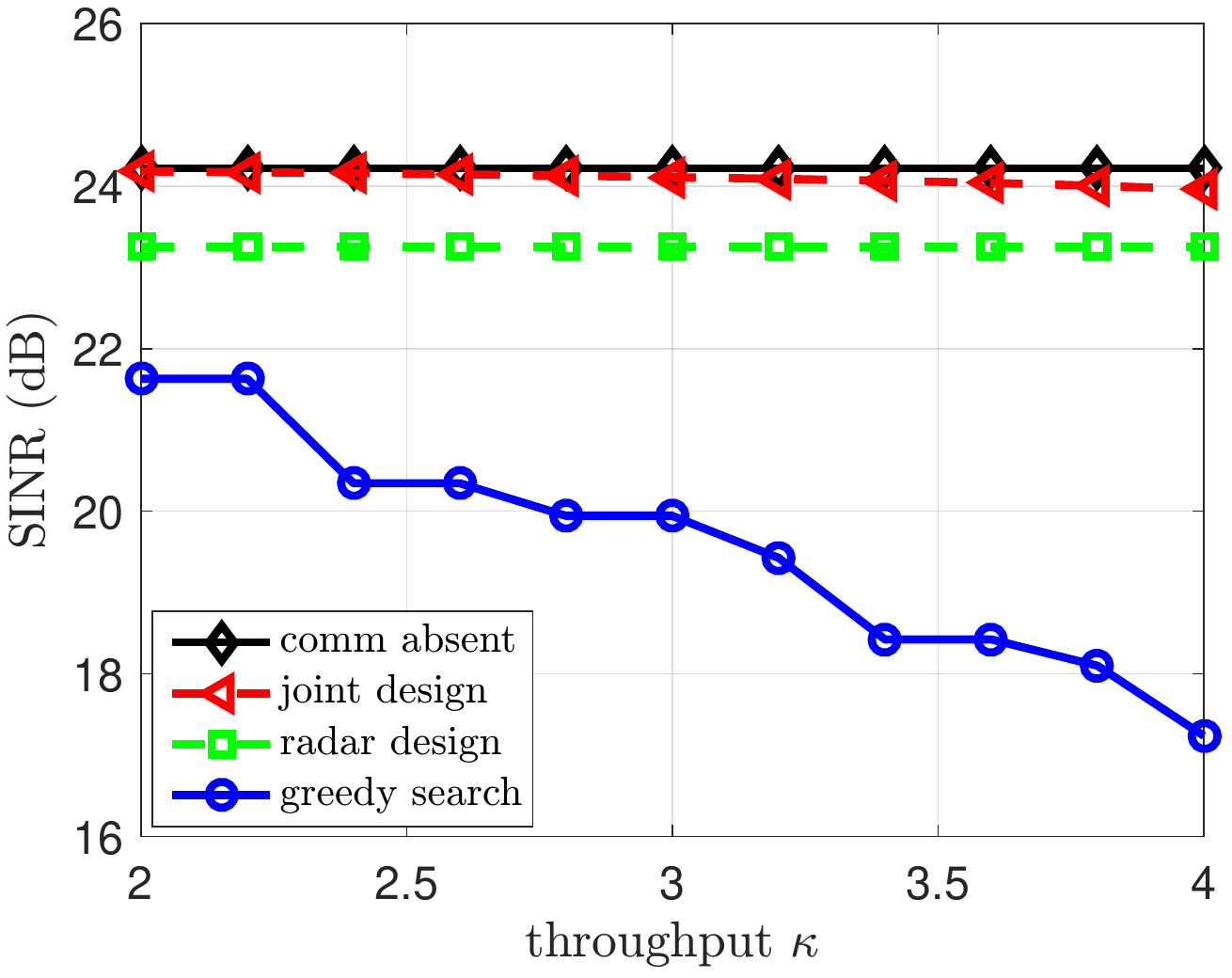}\\ 
(a)\\
\includegraphics[width=3in]{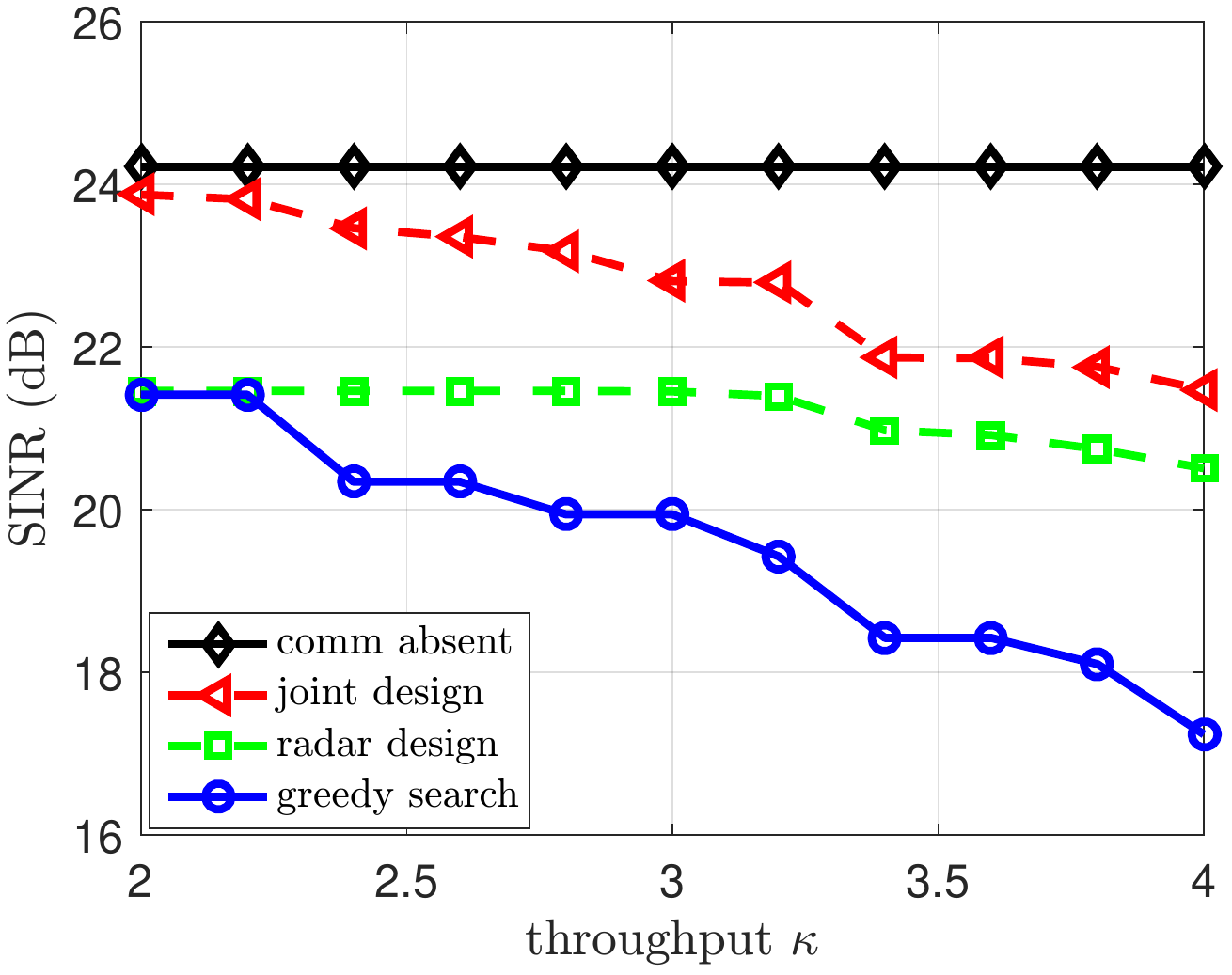}
\\
(b)
\et
\caption{Output SINR versus the communication service constraint $\kappa$. (a) Weak cross interference (Case 1); (b) strong cross interference (Case 2).}
\label{fig:subfigureCap}
\end{figure}

Next, we evaluate the effects of the communication throughput constraint. Fig.~\ref{fig:subfigureCap} shows the output SINR versus $\kappa$, where $P_\text{r}=P_\text{c}=600$. It can be seen that as the communication throughput constraint increases, the output SINR of all methods except for the comm absent degrades. This is because the communication needs to increase its transmission power to meet the increasing throughput constraint, which causes stronger interference to the radar system. Note, however, that the degradation of the joint design and unilateral design is considerably smaller in Case 1 than that in Case 2.

\begin{figure}[!hbt]
\centering
\bt{cc}
\includegraphics[width=3in]{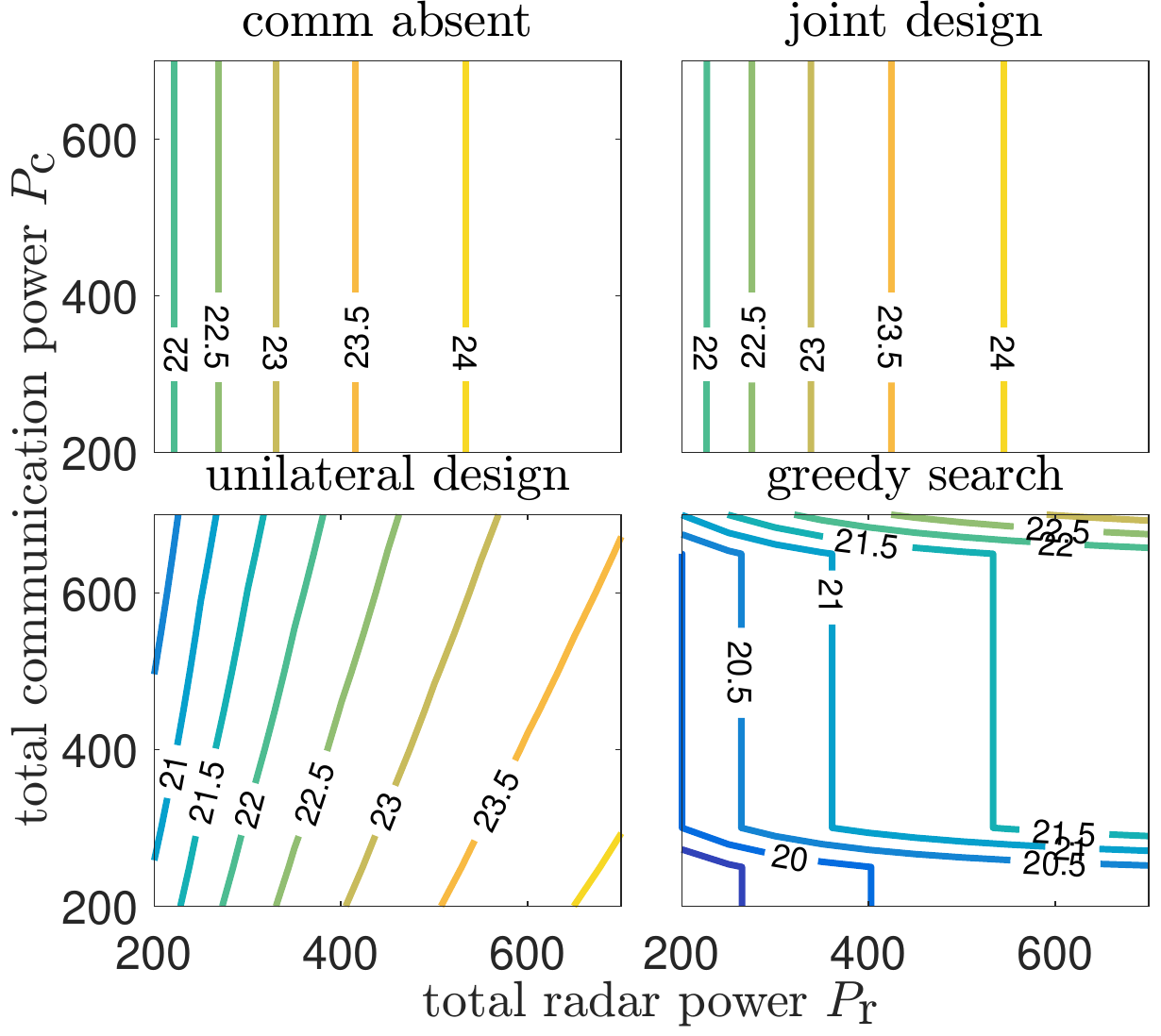}\\ 
(a)\\
\includegraphics[width=3in]{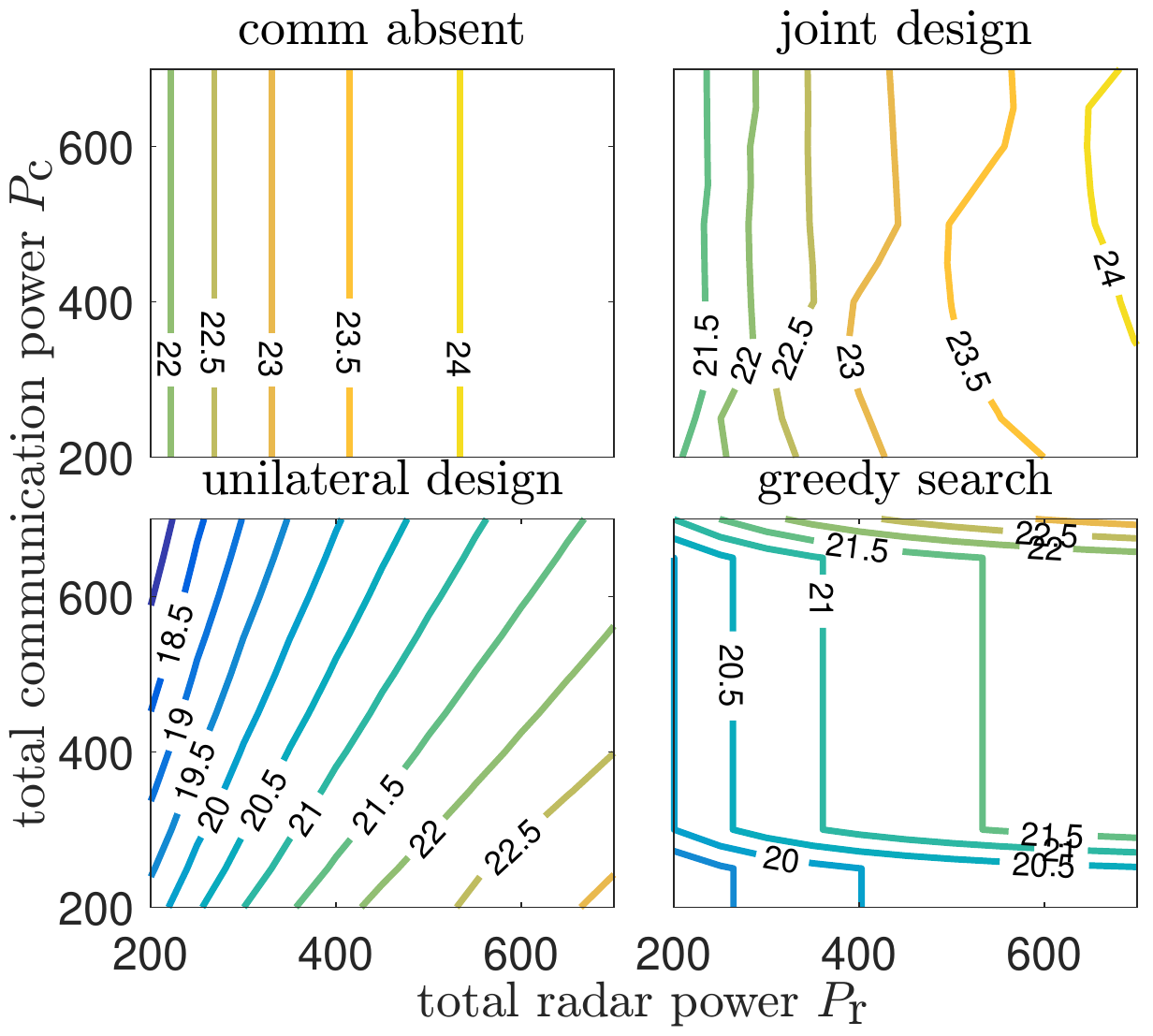}
\\
(b)
\et
\caption{Contour plot of the output SINR versus the total radar power $P_\text{r}$ and total communication power $P_\text{c}$. (a) Case 1; (b) Case 2.}
\label{fig:contourplot}
\end{figure}

Fig.~\ref{fig:contourplot} depicts the contour plot of the output SINR versus the total transmission power $P_\text{r}$ and $P_\text{c}$. Each plot contains the isolines of the output SINR with a stepsize of 60. For the comm absent design, the contour lines are vertical, as its output SINR only depends on the total radar transmission power. The contour plot of the joint design is almost identical to that of the comm absent in Case 1 as indicated by Fig.~\ref{fig:contourplot} (a). This is because when the cross interference is weak, the impact from one system to the other is limited. On the other hand, the greedy search design has the worst performance since it requires the most radar and communication transmission power to achieve the same output SINR.

Fig.~\ref{fig:contourplot} (b) shows that the comm absent and greedy search in Case 2 share the same performance trend as those in Fig.~\ref{fig:contourplot} (a) since they are independent of cross interference. On the other hand, both the joint design and unilateral design are observed to degrade in Case 2, although the latter exhibits a larger performance degradation.

\begin{figure}[!hbt]
\centering
\includegraphics[width=3in]{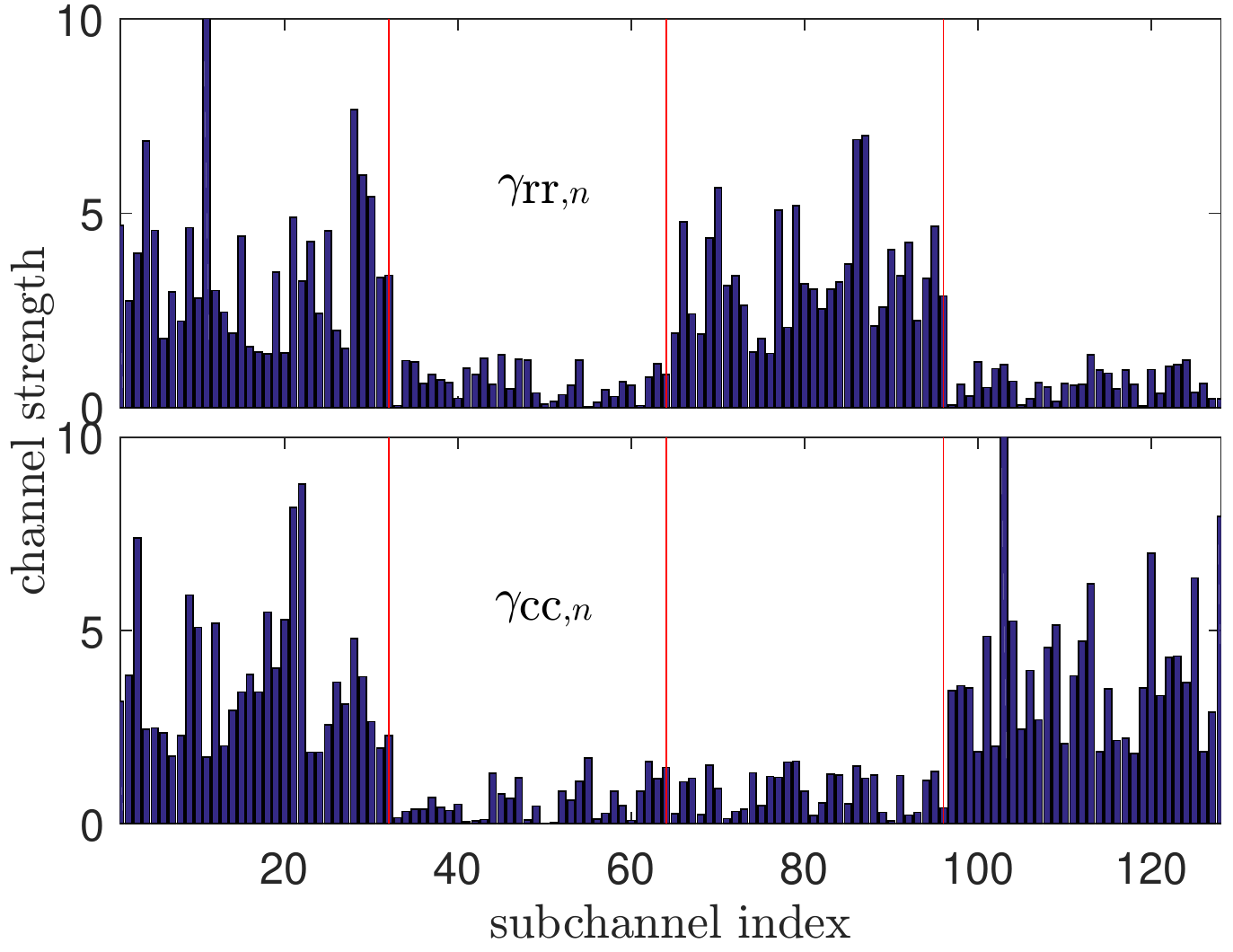}
\caption{The channel strength of $\gamma_{\text{rr},n}$ and $\gamma_{\text{cc},n}$.}
\label{fig:channel}
\end{figure}


To offer further insight, we look into the specific power allocation provided by different designs. We assume the multicarrier systems employ $N=128$ subcarriers divided into four groups with each consisting of 32 subcarriers. The normalized INRs for the cross interference $\eta_{\text{rc},n}$ and $\eta_{\text{cr},n}$ are fixed as $\eta_{\text{rc},n}=\eta_{\text{cr},n}=0.01$, $n=1,\dots,128$. The normalized CNR is $\eta_{rr,n}=0.05$. The desired channel strength $\gamma_{\text{rr},n}$ and $\gamma_{\text{cc},n}$ are depicted in Fig.~\ref{fig:channel}, which show the first group of subcarriers ($n=1,\dots,32$) is good for both radar and communication, the second group ($n=33,\dots,64$) bad for both systems, the third group ($n=65,\dots,96$) good for radar but bad for the communication system, and the fourth group ($n=97,\dots,128$) is the opposite of the third group. The other parameters are $P_\text{r}=P_\text{c}=600$ and $\kappa=2.5$ bits/s/Hz.

\begin{figure}[!hbt]
\centering
\bt{cc}
\includegraphics[width=3in]{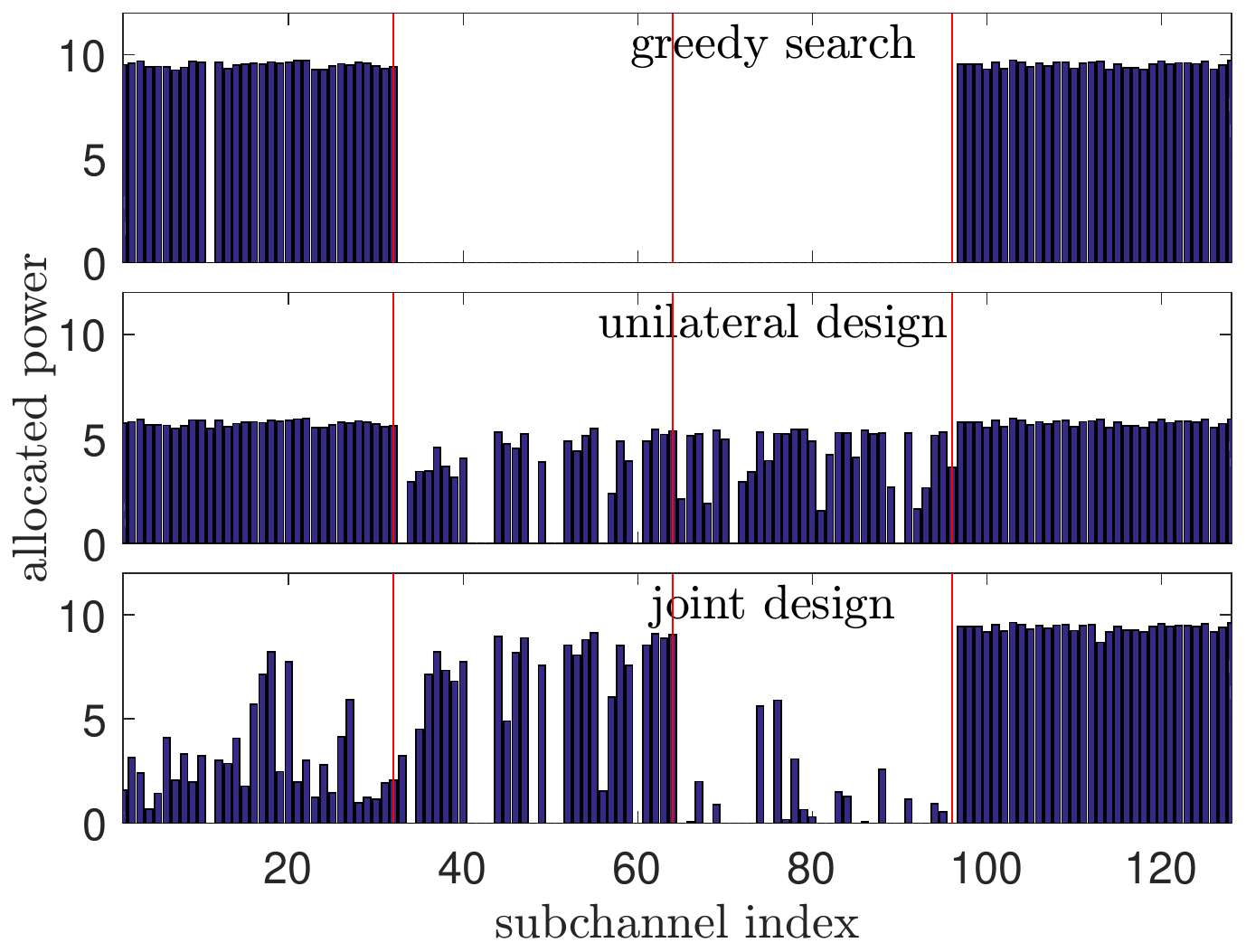}\\
(a)\\ 
\includegraphics[width=3in]{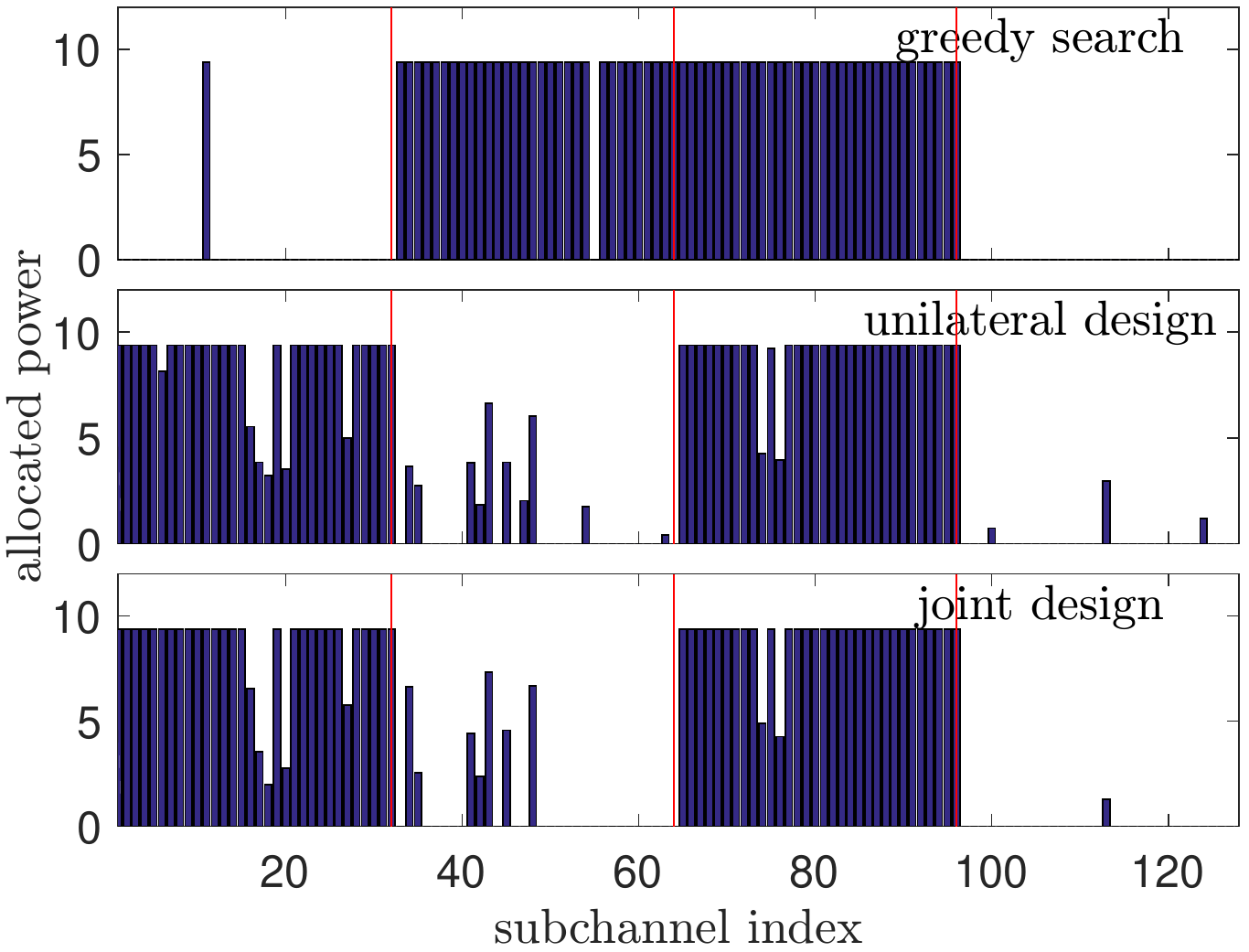}
\\
(b)
\et
\caption{Power allocation for $P_\text{r}=P_\text{c}=600$ and $\kappa=2.5$ bits/s/Hz. (a) Communication power allocation; (b) radar power allocation.}
\label{fig:subfigurepower}
\end{figure}

The specific subcarrier power allocation results are shown in Fig.~\ref{fig:subfigurepower}, where the resulting SINR obtained by greedy search, unilateral design, and joint design are 28.4, 30.2 dB, and 31.3 dB, respectively. It is seen from Fig.~\ref{fig:subfigurepower} (a) that the greedy search design assigns to the communication system from its best subcarriers ($n=1,\dots32$ and $n=99,\dots,128$) until the throughput constraint is satisfied, whereas the radar employs the rest subcarrier to maximize its SINR as shown in Fig.~\ref{fig:subfigurepower} (b). For the unilateral design, the communication system first utilizes waterfilling to allocate its power [cf. \eqref{equ:P_u}], and then the radar maximizes its output SINR based on \eqref{equ:P4}. Interestingly, it is observed that the joint design reduces the communication power on the first and third groups of subcarriers to lower its interference to the radar and at the same time, increases the communication power on groups 2 and 4. This leads to an improved SINR for the radar system.

\section{Conclusion}
\label{sec:conclusion}
Power allocation based spectrum sharing between multicarrier radar and communication systems was considered by maximizing the radar output SINR while meeting a communication throughput requirement along with total/peak power constraints. A joint design as well as a unilateral design were proposed to tackle the coexistence problem. Through suitable reformulation, the nonconvex joint design was solved by a computationally efficient non-alternating method, while the unilateral design was solved by a Taylor expansion based iterative convex relaxation procedure. Simulation results validated the effectiveness of the proposed spectrum sharing methods over the subcarrier-allocation based GS scheme.


\bibliographystyle{IEEEtran}
\bibliography{MC20}
\end{document}